\documentclass[12pt]{article}

\usepackage{times}

\usepackage{graphicx}
\usepackage{url}

\newenvironment{sciabstract}{%
\begin{quote} \bf}
{\end{quote}}

\title{ Dynamics of epidemic diseases without guaranteed immunity}

\author
{Kurt Langfeld\\
\\
\normalsize{School of Mathematics, University of Leeds, LeedsLS2 9JT, UK }
\\
\normalsize{k.langfeld@leeds.ac.uk}
}

\date{31 July 2020 }

\begin{document} 

% Double-space the manuscript.

\baselineskip24pt

% Make the title.

\maketitle 

\begin{sciabstract}
  The global SARS-CoV-2 pandemic suggests a novel type of disease
  spread dynamics.  WHO states that there is currently no evidence that
  people who have recovered from COVID-19 and have antibodies are
  immune from a second infection~\cite{WHO}. Conventional mathematical
  models consider cases for which a recovered individual either becomes
  susceptible again or develops an immunity. Here, we study the case
  where infected agents  recover and only develop immunity if they are
  continuously infected for some time. Otherwise, they become
  susceptible again. We show that field theory bounds the peak of the
  infectious rate. Consequently, the theory's phases characterise the
  disease dynamics: (i) a pandemic phase and  
  (ii) a response regime. The model excellently describes the epidemic
  spread of   the SARS-CoV-2 outbreak in the city of Wuhan, China. We
  find that  only 30\% of the recovered agents have developed an
  immunity. We anticipate our paper to influence the decision making
  upon balancing the economic impact and the pandemic impact on
  society. As long as disease controlling measures keep the disease
  dynamics in the ``response regime'', a pandemic escalation ('second
  wave') is ruled out. 
\end{sciabstract}

\paragraph*{Introduction}

The rapid spread of a disease across a particular region or regions
(epidemic) or the global outbreak of a disease
(pandemic)~\cite{porta2008dictionary} can have
a detrimental effect on health systems, on local and global economies
including the financial markets and the socio-economic
interactions, 
ranging from the city to the international level. Measures to reduce
the pandemic spread include curtailing
interactions between infected and uninfected parts of the population, reducing
infectiousness or the susceptibility of members of the
public~\cite{Ferguson2005}.  The two major strategies governments use
to handle an 
outbreak are  to slow down an outbreak (mitigation) or to  interrupt
the disease spread (suppression).  Since each of those interventions
bears itself significant risks for 
the societal and economic well-being, it is crucial to understand
the effectiveness of these strategies (or any hybrid of them).

Mathematical methods provide essential input for governmental decision
making that aims at controlling the outbreak. Among those are statistical
methods~\cite{Unkel2012,Becker1999}, deterministic state-space
models~\cite{Brauer2019} with its prototype developed by Kermack and
McKendrick~\cite{Kermack1927}, and a variety of complex network
models, e.g.,~\cite{Hwang2005,SHIRLEY2005287}.
The different mathematical approaches have different objectives:  A
significant 
application of the statistical methods frequently aims at the early detection
of disease outbreaks ~\cite{Unkel2012}, while modelling either tries
to develop a model as realistic as possible for a given outbreak or to
design a simplistic model, which, however, reveals some universal truth
about the outbreak dynamics.

In the simplest version, the so-called compartmental
models~\cite{Kermack1927,Hethcote2006} consider the fraction of the population
which is either susceptible (S), infected (I) or removed (R) from the
disease network. Coupled differential equations capture the dynamics of
the disease
that determine the time dependence of S, I and
R. Extensions add more compartments to the SIR model such as (E)
exposed. For example, such an SEIR model was used in~\cite{Lekone2006} for a
description of the Ebola outbreak in the Democratic Republic of Congo
in 1995. Compartmental models have been applied to describe the recent
SARS-CoV-2
outbreaks~\cite{Giordano2020,KRISHNA2020375,TAGLIAZUCCHI2020109923,LIN2020211,Anastassopoulou2020,wu2020}. For
example, 
the elaborate model from Giordano at al.~uses a total of 8
compartments - susceptible (S), infected (I), diagnosed (D), ailing
(A), recognized (R), threatened (T), healed (H) and extinct (E) - to
describe the COVID-19 epidemic in Italy. Compartmental models have
been extended in order to capture
stochastically unknown influences, such as changing
behaviours~\cite{10.1093/biostatistics/kxs052}. Such models were
recently used to analyse the COVID-19
outbreak in Wuhan~\cite{KUCHARSKI2020553}. 

Compartmental models address \texttt{global} quantities such as the
fraction of susceptible individuals S and assume that heuristic rate
equations can describe the disease dynamics. In
cases of a strongly inhomogeneous (social) network, e.g. taking into
account different population densities, the above assumption seems not
always be justified. In these cases, spatial disease spread patterns
can be described by a stochastic network model with
Monte-Carlo simulations a common choice for the simulation.

\medskip
In this paper, we consider a disease dynamics for which the duration
(severity) of the illness depends on the amount of exposure. Using an
elementary (social) network, we are looking for universal mechanisms
describing a pandemic spread. We will reveal a connection to statistical field
theory, enabling us to characterise an outbreak with the tools of
critical phenomena. We will discuss the impact of the findings on
policies to curb an outbreak and will draw conclusions from the
recent Covid19 outbreak in Hubei, China.

\paragraph*{Model basics:}

We assume that each individual has two states $u$
with $u=0$ - susceptible and $u=1$ - infected. Each individual
interacts with four 'neighbours' of the social network. The disease
spread is described as a stochastic process. At each time step (say
'day'), the probability that an individual gets infected (or recovers)
depends on the status of the neighbours in the social network. Here, 
we only study the simple case of a homogeneous network with four
neighbours for each site. We also consider periodic boundary
conditions to minimise edge effects.

\paragraph*{Immunity:}

We study two closely related scenarios.
\begin{enumerate}
\item[(i)]There is no immunity. Every individual can be reinfected
  and can recover only to be susceptible again. 

\item[(ii)] Individuals can be reinfected and recover. Only if individuals
  stay infected for $\tau $ consecutive days, they are considered
  \textit{immune}. 

\end{enumerate} 
In case (ii), the sites of immune individuals are removed from the disease network.

\paragraph*{Disease dynamics:}
If $x$ is a site of the disease
network, at every time step the state $u_x \in \{0,1\}$ is randomly chosen with
probability 
\begin{equation}
P(u_x) \, = \, \frac{1}{{\cal N}_x} \, \exp \left\{ (4 \beta  \, n_x
  \, + \, 2 h ) \, u_x\right\} , 
\, \; \; \;  n_x \, = \, \sum _{y \in \langle xy \rangle } u_y \; , 
\label{eq:0}
\end{equation} 
where $\langle xy \rangle$ is an elementary link on the lattice
joining sites $x$ and $y$ and, hence, $n$ is the number of infected
neighbours, and ${\cal N }_x  =1 \, + \, \exp \left\{ 4 \beta \, n_x + 2 h
\right\} $ is the normalisation. The parameter
$\beta $ describes the {\it contagiousness } 
of the disease. The parameter $h$ is linked to the probability to contract the disease
from outside the network. In fact, if no-one of the network is
infected ($n_x=0, \, \forall x$), the probability $p$ that any
individual contracts the diseases, is connected to $h$ by 
$$
 p \; = \; \frac { \exp \{2h\} }{ 1 + \exp\{ 2h \} } \; . 
$$
If the lattice contains $N$ individuals (i.e.,
sites), one time step is said to be completed if we have considered 
$N$ randomly chosen sites for the update. 

\paragraph*{The pandemic spread as a critical phenomenon:}

\begin{figure}
  \includegraphics[height=6.5cm]{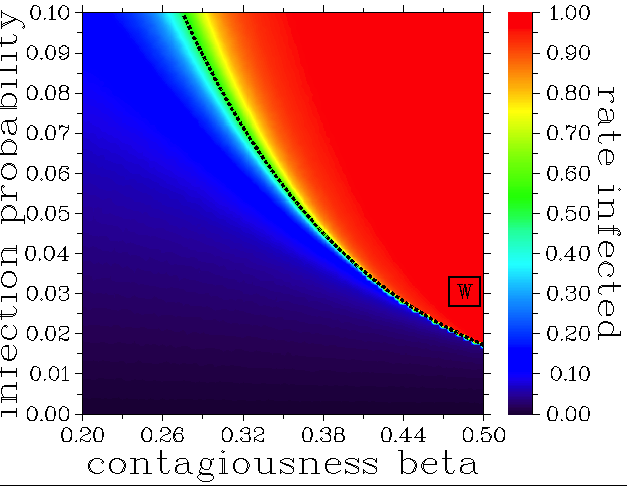} \hspace{0.5cm}
  \includegraphics[height=6.5cm]{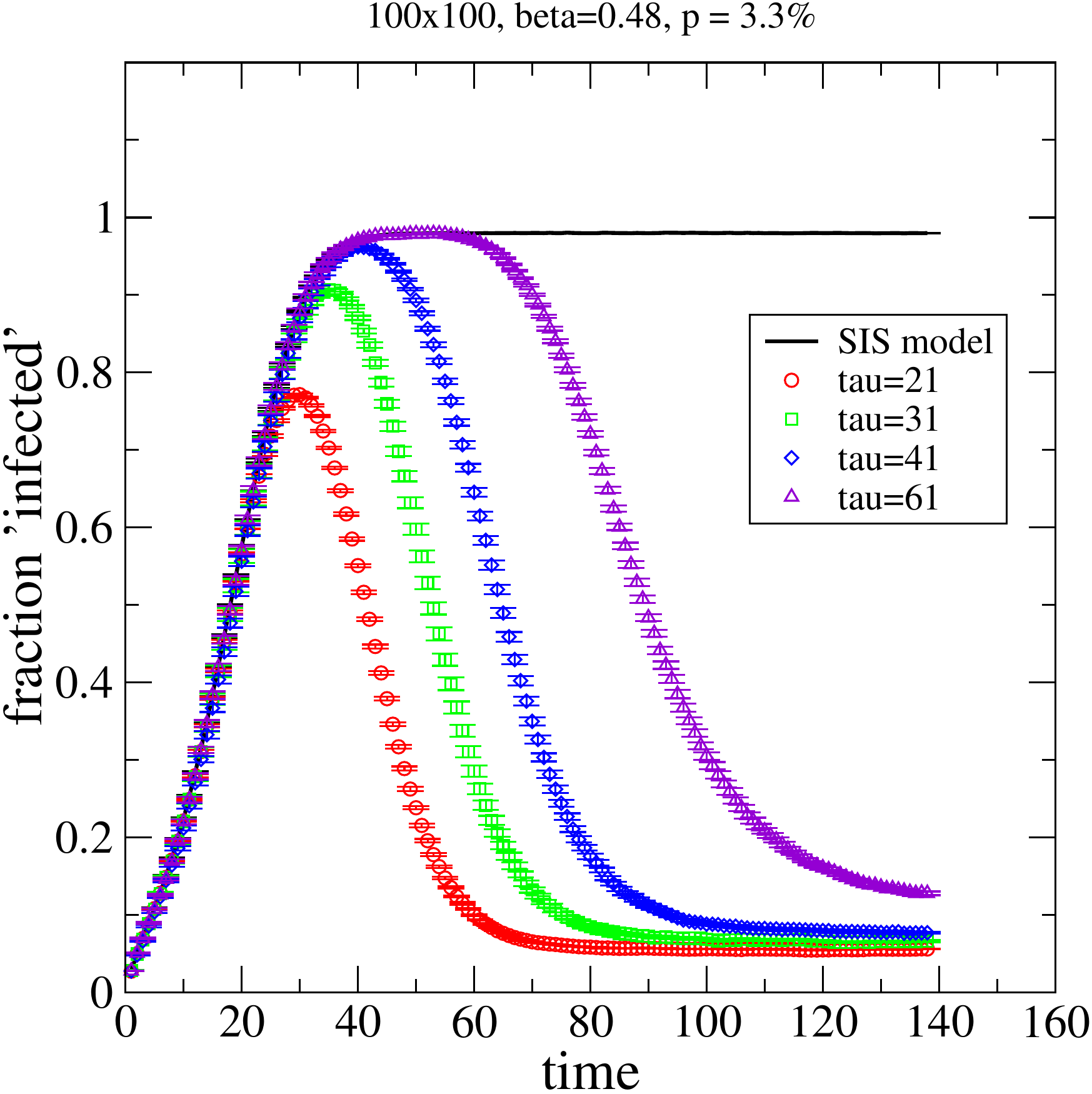}
  \caption{\label{fig:1} Left: Asymptotic state of the scenario (i):
    average infection rate as a function of contagiousness $\beta $ and
    see infection probability $p$. Also shown is the critical line $p(\beta)$ in
    (\ref{eq:1}) (black dashed line). ``W'' indicates the parameter set
    consistent with COVID-19 outbreak in Wuhan. Right: Dynamics of the
    rate of infected (red bars)     compared with the asymptotic value
    of the field theory, which   bounds the maximum rate of infected.}
\end{figure}
Scenario (ii) show the typical time evolution of an epidemic with the
infections rate approaching zero for large times due to agents
recovering and an increasing number being immune. By contrast,
scenario (i) has an asymptotic state 
independent from the initial state and described by statistical field
theory. After the change of variable $z_x=2u_x-1$, the asymptotic
state is described by the partition function of the Ising
model~\cite{Ising1925,friedli2017}: 
$$
Z \; = \; \sum_{\{z_x=\pm 1\}} \exp \Bigl\{ \beta \sum _{\langle
  xy\rangle } z_x \, z_y + H \sum _x z_x \Bigr\} 
$$
with $H = h + 4 \beta $, which is the well-known partition function
for Ising spins $z$ in an external magnetic field $H$. The disease dynamics of
scenario (i) corresponds to a Markov chain of local updates in the
Ising model with Markov time identified as real time. 
\begin{equation}
H \; = \; 0  , \; \; \;  h(\beta ) \; = \; - 4 \, \beta   , \; \; \;
p(\beta) \; = \; \frac{1}{e^{8 \beta } +1 } . 
\label{eq:1}
\end{equation}
For a vanishing external field $H$, the model shows a critical
behaviour with a phase transition at $\beta = \beta _c  = \ln ( 1 +
\sqrt{2} )/2 \approx 0.44$. In the ordered phase for $\beta > \beta
_c$, a small seed probability $p>0$ triggers an infection rate close to 100 \% of
the population. The model is in the 'pandemic' phase. For $\beta <
\beta _c $, the model is in the 'response' phase, i.e., the infection
rate is in repose to the seed probability $p$, but no outbreak occurs. The asymptotic
infection rate can be calculated using Markov Monte-Carlo methods. We
used a modified Swendsen-Wang cluster algorithm, which performs near
the phase transition~\cite{swendsen1989}. Our numerical findings are
summarised in figure~\ref{fig:1}, left panel. Curve~\ref{eq:1}
clearly separates both phases - the pandemic phase and the response
regime. Note that the dynamics (\ref{eq:0}) corresponds to the standard
heat-bath update~\cite{kendall2005}. Starting from a healthy
population ($u_x=0, \, \forall x$), it 
takes the 'thermalisation' time $t_\mathrm{th}$ that the daily infection rate starts
fluctuating around its asymptotic value. 

\begin{figure}
  \includegraphics[height=6cm]{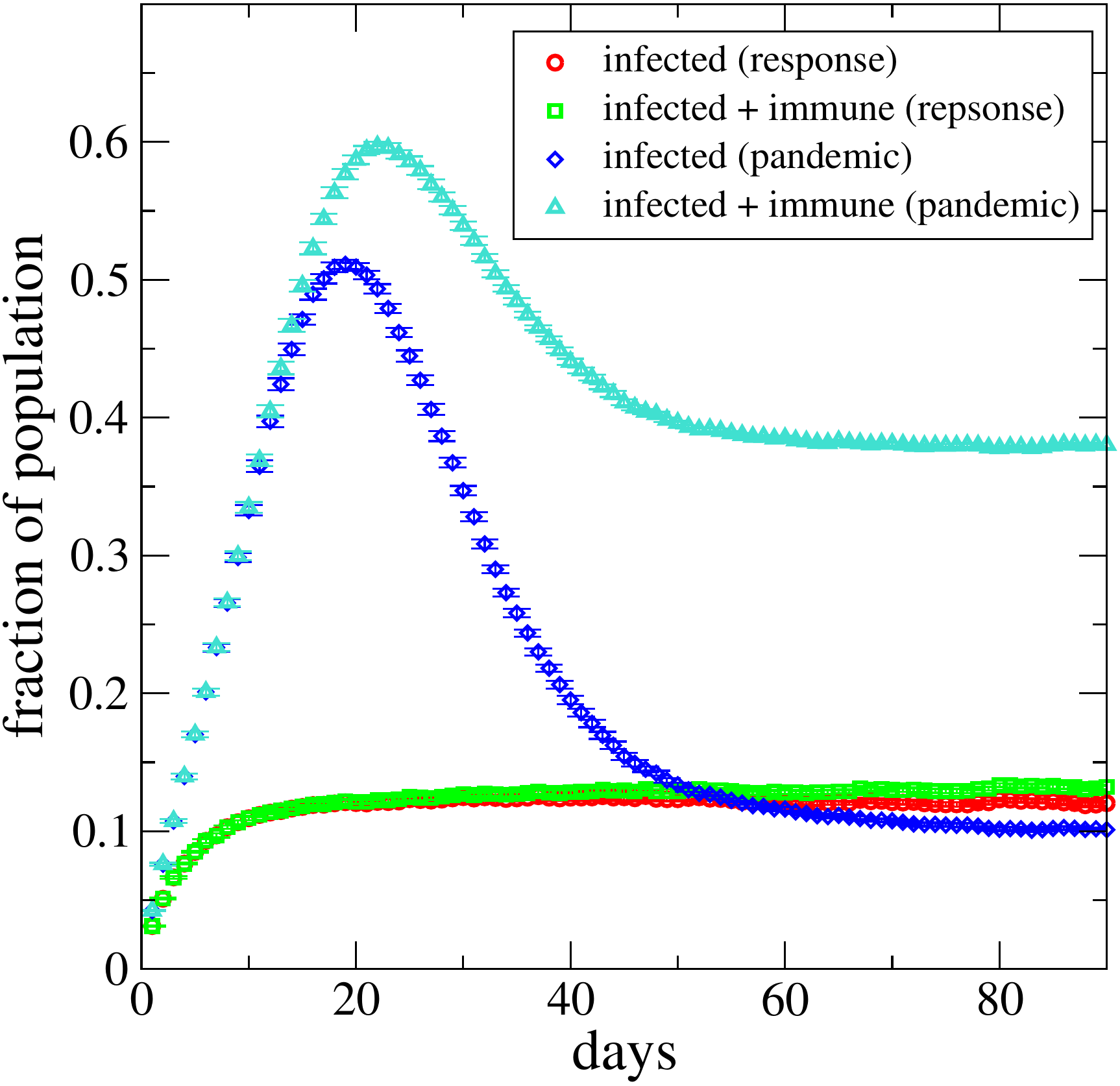} \hspace{0.5cm}
  \includegraphics[height=6cm]{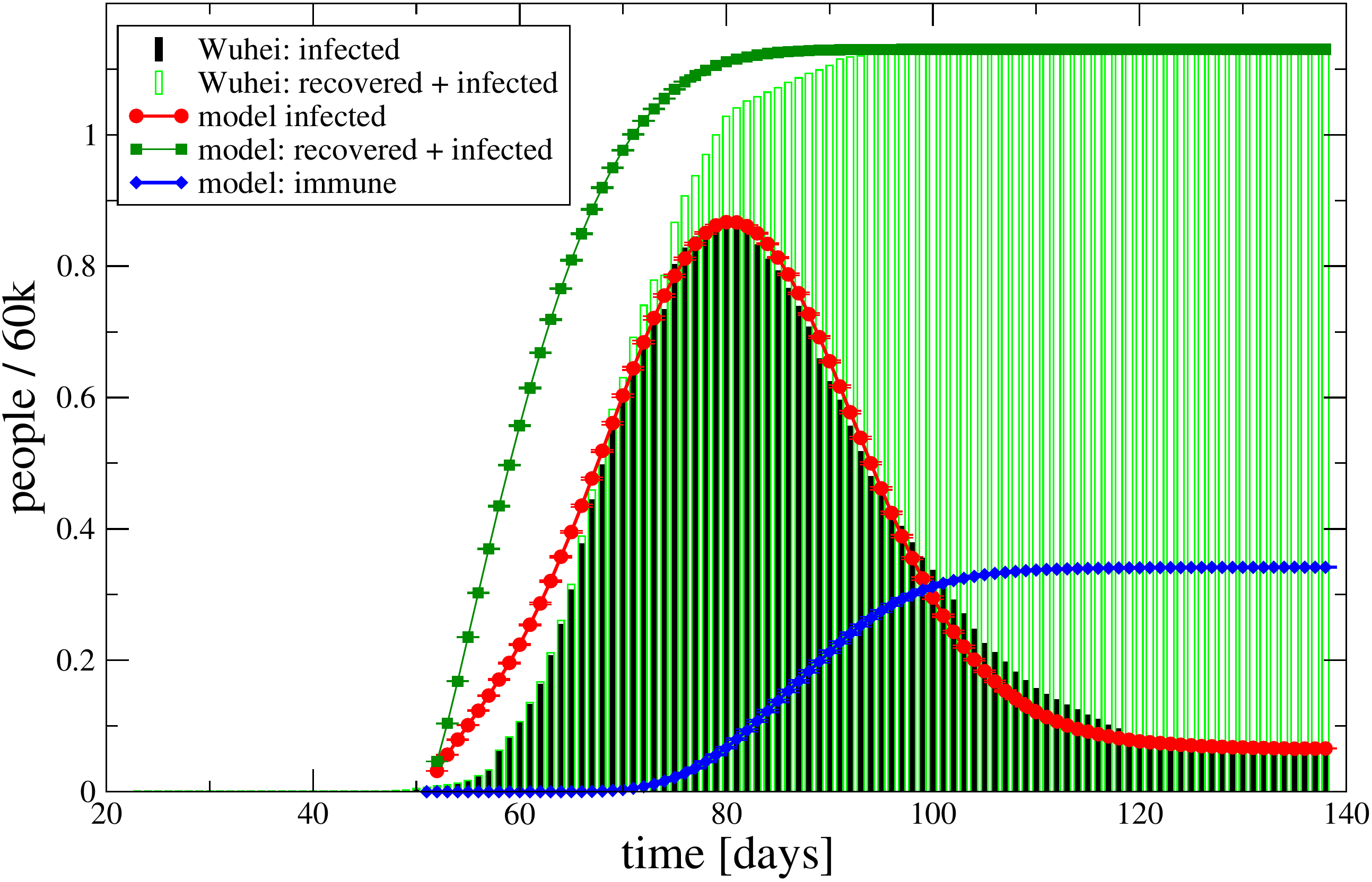}
  \caption{\label{fig:3} Left: Scenario (ii): rate of infected (red bars) and rate of
    infected $+$ immune (green bars) for two sets of model parameters:
    pandemic
    phase ($p=0.05, \, \beta = 0.41$) and response regime
    ( $p=0.04, \, \beta = 0.38$) both for $\tau =11$. 
    Right:  Comparison of scenario (ii) results with
    actual data from the COVID-19 outbreak in Hubei province.}
\end{figure}
\paragraph*{Immunity:}
Let us now study scenario (ii), where individuals
can develop immunity if they are infected for $\tau $
consecutive days. For $\tau> t_\mathrm{th}$, the peak infection rate
is that of the asymptotic 
state of the corresponding model (i) and, hence, inherits the
classification 'pandemic' or 'response' phase. This is illustrated in
figure~\ref{fig:1}, right panel, for the pandemic phase for several
values of $\tau $. Figure~\ref{fig:3} 
illustrates the vastly different behaviour of the disease spread in
the pandemic phase ($\beta = 0.41, \, p =5\%$) and in the response
regime ($\beta = 0.38, \, p = 4\%$). Results are for a $N=100 \times
100$ network and $\tau = 11$. Note that the curve for
'infected+immune' ('triangle' symbol) in the pandemic phase is
{\it not} monotonically increasing with time since the infected individuals
can return to 'susceptible' state, i.e., not every
infected individual becomes immune. Note that in the
response regime ('circle' and 'square' symbol), the 'pandemic' peak is absent
altogether. However, on the downside, the so-called 'herd immunity'
slowly develops over time. 

\paragraph*{Comparison with data:}
We stress that the model assumption of a homogeneous (social) network
with 'four neighbours' is unrealistic. The knowledge of the
underlying disease network is essential to make quantitative
predictions for e.g. the critical value $\beta _c$ of the
contagiousness. Here, we adopt a different approach: we assume that
qualitative time evolution of bulk quantities such as the
fraction of infected individuals is within the grasp of model scenario
(ii) and use those as fit functions to determine the model 
parameters such as $\beta $, $p$ and $\tau $ by comparison with actual
data.

For this study, we used data from the COVID-19 outbreak in 2020 in the Hubei
province in China~\cite{Yu2020}. The data of the number of infected
individuals show a jump at day 73 (on the arbitrary time scale) by
40\%, which is due to a change in reporting. We assume that the same
'under-reporting' has occurred in the days before, and have corrected
the data by multiplying the number of infected (and infected+recovered)
by a factor $1.4$ for times $t\le 73$. Let $D(t, \tau, \beta, p)$ be the
fraction of the population of infected individuals as a function of
time $t$ and depending on the parameters $\tau $ (time to develop
immunity), $\beta $ (contagiousness) and $p$ seed probability to get
infected. We have calculated $D(t, \tau, \beta, p)$ using a
$N=250\times 250$ lattice. We checked that the result is independent
of the lattice size in the percentage range for the parameters relevant
in this study. If $D_\mathrm{wuhei} (t)$ quantifies the measured values for
the number of infected in the Hubei outbreak, we want to approximate
these data, i.e.,
$$
D_\mathrm{wuhei} (t) \; \approx \; N_\mathrm{pop} \, D(t - t_s, \tau, \beta, p) 
$$
with a suitable choice of the parameter $N_\mathrm{pop}$, $ t_s$, $\beta $ and
$p$. Since the offset of the time axis in the Hubei data is arbitrary,
we have chosen the shift $t_s$ such that the peaks of simulated data and
measures data coincide. All other parameters are treated as fit
parameters. Altogether, we find a good agreement with the data for: 
$$
N_\mathrm{pop}\approx 68k, \; \; \; t_s \approx 50, \; \; \; \tau \approx 21, \;
\; \; \beta \approx 0.48,
\; \; \; p \approx 3.3\% \; . 
$$
The model data overshoot the data in the early days of the epidemic
spread, which could be related to underreporting due to limited 
testing capabilities. It is interesting to observe that the curve of
the infection rate is asymmetric: the slope of the raise at the
beginning is larger than the slope of the decline
after the maximum. Also, the number of infected seem to level off at a
non-zero value. In the present model, this is explained as follows:
with more agents being immune, it is harder for susceptible agents to be 
continuously infected  for time greater or equal $\tau $ and, thus, to
develop immunity. We also find that only about 30\% of the infected
(and recovered) develop an immunity.

\medskip \noindent
\paragraph*{Conclusions and interpretations:}
A new type of stochastic disease model is proposed: agents can recover
from an infection and are susceptible again. They only develop 
immunity if their infection lasts
longer than a characteristic time $\tau $. For $\tau \to \infty $, the
infection rate is described by statistical field theory. For finite
$\tau $, the infection rate of the field theory provides an upper
bound of the infection rate of the dynamical model. This opens up the
possibility to characterise the disease dynamics in the light of critical
phenomena of the underlying field theory: a pandemic spread
corresponds to the ordered phase of the field theory, and the critical
value for the contagiousness is 
that for the phase transition. The disease is in controllable
response mode if the corresponding field theory is in the disordered
phase.

\medskip
Quantitative results, reported here, are derived with an unrealistic
homogeneous disease network for which each agent interacts with four
neighbours. Nevertheless, we find that the Covid19 data of the Hubei
outbreak are well represented. For this case, we find that only $30\%$ of infected
develop an immunity.

\medskip
The heavy tail of the decline of the number of infected, which levels
off at non-zero values, is an inherent feature of the model and can be
traced back to the fact that agents can be reinfected. In a network with a sizeable
portion of immune agents, it is increasingly challenging to develop
immunity. If these model assumptions were underpinned by medical
investigations, achieving 'herd immunity' would be difficult. This 
should influence the decision to what extent efforts  focus on developing
a cure or a vaccine.

\section*{Acknowledgments}
I thank Lorenz von Smekal (Giessen) and Paul Martin (Leeds) for
helpful discussions.

\end{document}